\documentclass[Physsubmission, Phys]{SciPost}

\hypersetup{
    colorlinks,
    linkcolor={red!50!black},
    citecolor={blue!50!black},
    urlcolor={blue!80!black}
}

\usepackage[bitstream-charter]{mathdesign}
\DeclareSymbolFont{usualmathcal}{OMS}{cmsy}{m}{n}
\DeclareSymbolFontAlphabet{\mathcal}{usualmathcal}

\begin{document}
\begin{center}{\Large \textbf{
Lund jet plane for Higgs tagging\\
}}\end{center}

\begin{center}
Charanjit K. Khosa
\end{center}
\begin{center}
Dipartimento di Fisica, Universit\`a di Genova and INFN, Sezione di Genova, Via Dodecaneso 33, 16146, Italy
\\
charanjit.kaur@ge.infn.it
\end{center}

\begin{center}
\today
\end{center}

\definecolor{palegray}{gray}{0.95}
\begin{center}
\colorbox{palegray}{
  \begin{tabular}{rr}
  \begin{minipage}{0.1\textwidth}
    \includegraphics[width=30mm]{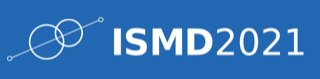}
  \end{minipage}
  &
  \begin{minipage}{0.75\textwidth}
    \begin{center}
    {\it 50th International Symposium on Multiparticle Dynamics}\\ {\it (ISMD2021)}\\
    {\it 12-16 July 2021} \\
    \doi{10.21468/SciPostPhysProc.?}\\
    \end{center}
  \end{minipage}
\end{tabular}
}
\end{center}

\section*{Abstract}
{\bf
We study the boosted Higgs tagging using the Lund jet plane. The convolutional neural network is used for the Lund images data set to classify hadronically decaying Higgs from the QCD background. We consider $H\to b \bar{b}$ and $H \to gg$ decay for moderate and high Higgs transverse momentum and compare the performance with the cut based approach using the jet color ring observable.  The approach using Lund plane images provides good tagging efficiency for all the cases. 
}


\section{Introduction}
\label{sec:intro}
Improved Higgs tagging will be beneficial for new physics searches as well as for precise standard model measurements at the Large Hadron Collider (LHC). Recent developments for Higgs tagging explored the potential of Machine Learning (ML) techniques along with novel observable construction, see e.g. \cite{Marzani:2019hun,Lin:2018cin}. Idea of using ML for jet tagging using low-level raw information (jet images) has been suggested a couple of years ago. We consider Lund jet images and  Convolutional Neural Networks (CNN)\footnote{CNNs provide very efficient image classification for different types of input data in High Energy Physics see e.g. \cite{Khosa:2019qgp, Khosa:2019kxd
}} for the signal and background classification for hadronically decaying boosted Higgs boson\cite{Khosa:2021cyk}. The Lund jet plane is a theory inspired jet representation which is used for jet tagging recently\cite{Dreyer:2018nbf, Dreyer:2020brq}\footnote{For experimental measurement of the Lund jet plane density see \cite{Aad:2020zcn, ALICE-PUBLIC-2021-002}.}. It uses the information of the radiation pattern inside the jet and is a log-log $(\ln \frac{1}{\Delta}, \ln \frac{k_t}{\text{GeV}})$ plane of the angle of the emission and the transverse momentum w.r.t the jet. We consider the primary Lund jet plane.  We assess the performance of this approach for both $H \to b \bar b$ and $H \to gg$ using two $p_T$ benchmarks (250 GeV and 550 GeV). We compare the classification accuracy with the standard cut based approach where one would consider observables. We consider a single observable named jet color ring, which by construction is an optimal observable for (two-prong) color singlet tagging\cite{Buckley:2020kdp}. It is defined as 
\begin{equation}
  \mathcal{O} = \frac{\Delta_{ka}^2+\Delta_{kb}^2}{\Delta_{ab}^2} \end{equation} 
where a and b are the primary (leading or tagged according to the decay products properties) subject, k is the remaining leading subject and $\Delta$ is the separation in the rapidity-azimuth plane.  

In our analysis, signal process we consider is $p p$ $\rightarrow$ $ZH$ where $Z \rightarrow \mu^+ \mu^-$ and $H \rightarrow$ $b \bar b$ or $gg$, and background processes are $Zbb$ and $Zjj$ for the $H \to b \bar b$ and $H \to g g$ analyses, respectively. We simulate events with generation level $p_T^{\mu\mu}$ cut of 200 and 500 GeV using \textsc{Madgraph 2.7.2}~\cite{Alwall:2014hca} at $\sqrt{s}=14$ TeV. Pseudo-rapidity cuts for leptons ($|\eta_l|<$2.5) and jets ($|\eta_j|<$5.0) are also imposed at the generation level. \textsc{Pythia 8.303}~\cite{Sjostrand:2006za} (with default Monash 2013 tune \cite{Skands:2014pea}) is used for the parton-shower and the hadronization.

We consider the particles excluding muons from Z decay and form radius R=1 jets using the anti-$k_t$ algorithm~\cite{antikt} implementation in \textsc{Fastjet 3.3.3}~\cite{fastjet}. We consider the leading jet with a $p_T$ cut of 250 GeV (or 550 GeV) and jet mass $110$ GeV $<m_J<$ 140 GeV for the analysis. We further need to identify the subjects inside the jet for the jet color ring. For that, we cluster the charged particles with $p_T$ >500 MeV using anti-$k_t$ algorithm (radius R=0.2) and form sub-jets. For sub-jets with $p_T>$ 5 GeV, we check if the  $\Delta<0.8$ with the leading large radius jet.  For the $H \to b \bar b$ analysis, we further identify these subjets as b-jets if $\Delta$ separation with the large radius jet is less than 0.2. Analysis efficiency is different for the signal and background processes as well as for different $p_T$ benchmarks. A large sample of events is generated such that after selections we get at-least 100K events for all the data sets. 
We re-cluster the constituents of the large radius jet using C/A algorithm, and consider the declustering history of this jet to get the primary Lund jet plane. We consider 25 by 25 pixels for the Lund jet images. For the CNN training, the data set of 200 K (equal proportions) signal and background events is divided into three sets: training set, validation set and test set in 60:20:20 ratios. 

\begin{figure}
\centering
\includegraphics[scale=0.47]{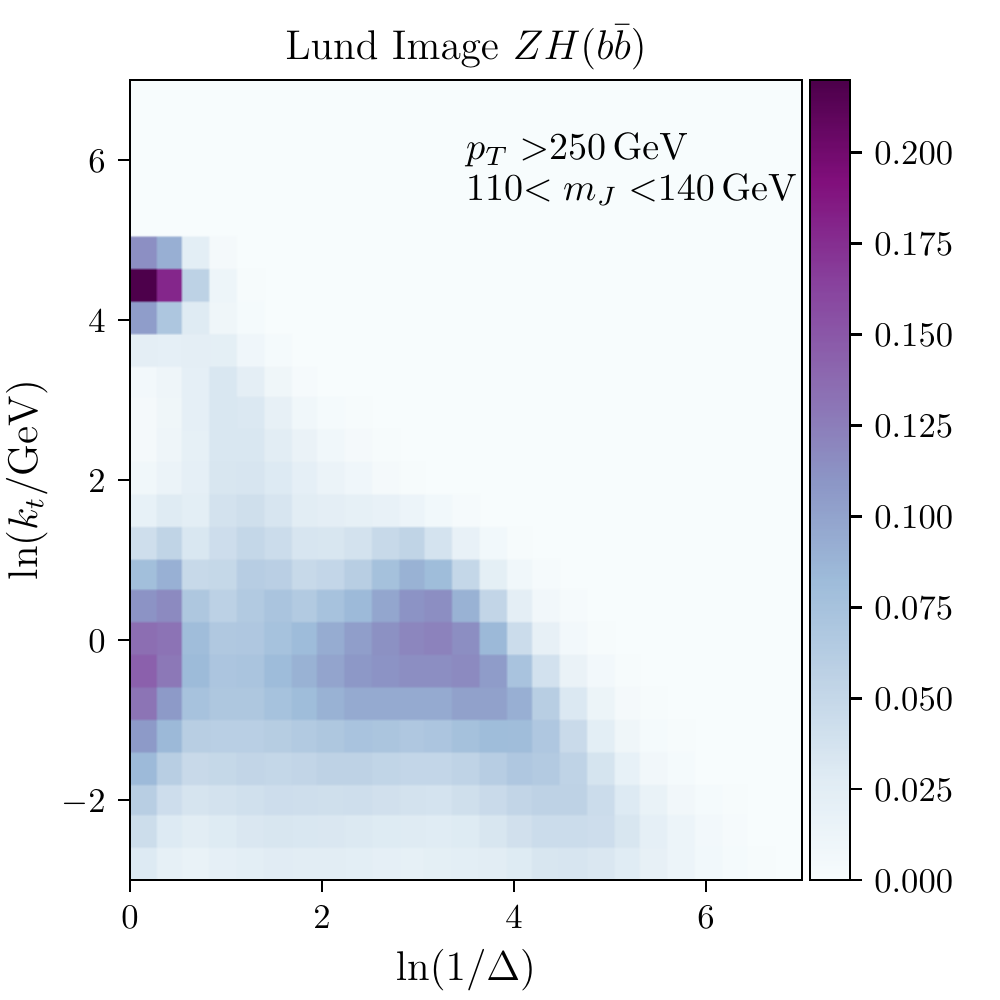}
\includegraphics[scale=0.47]{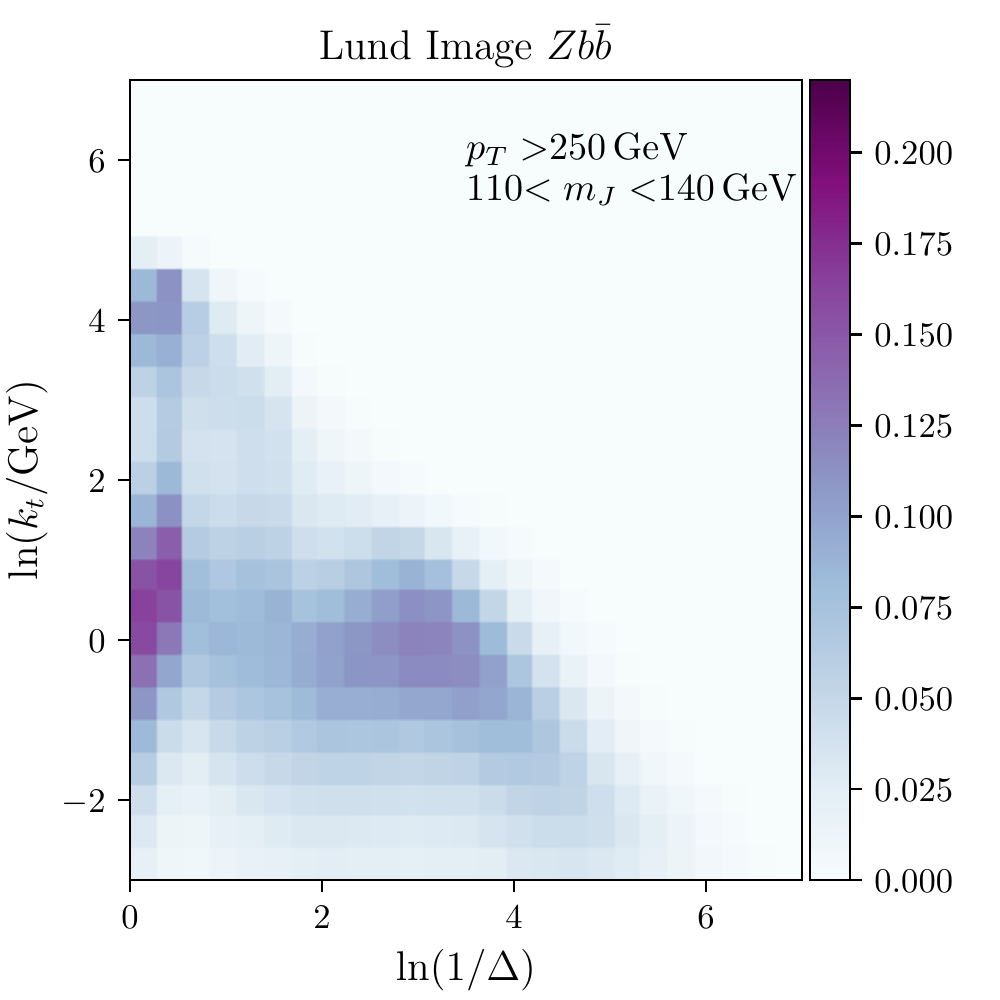}
\includegraphics[scale=0.47]{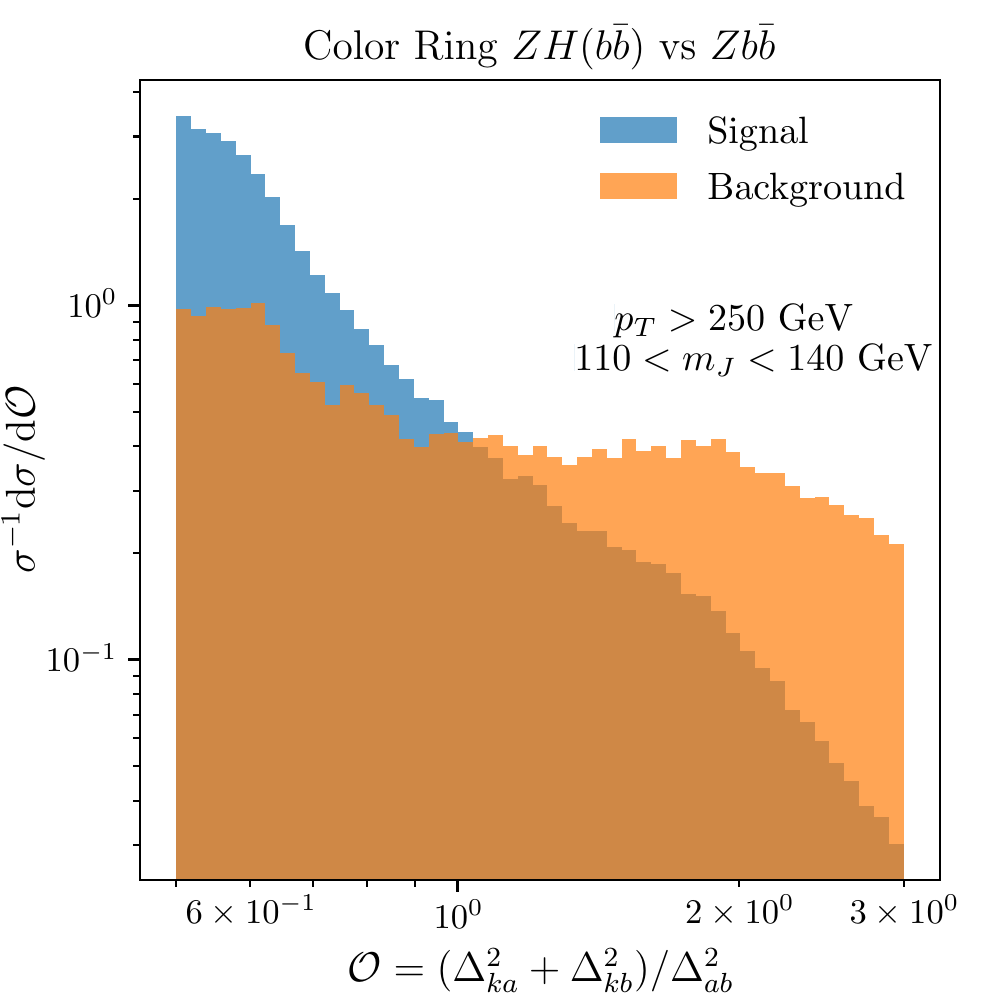}
\includegraphics[scale=0.47]{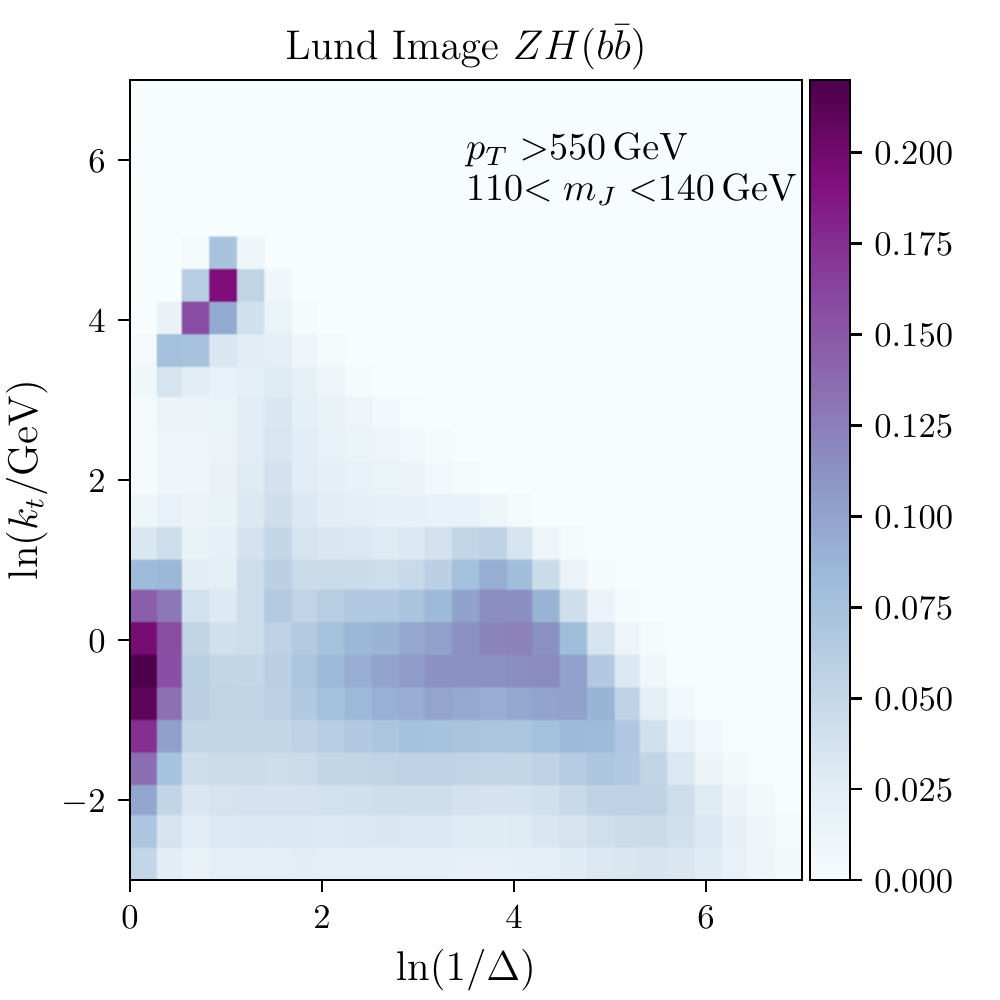}
\includegraphics[scale=0.47]{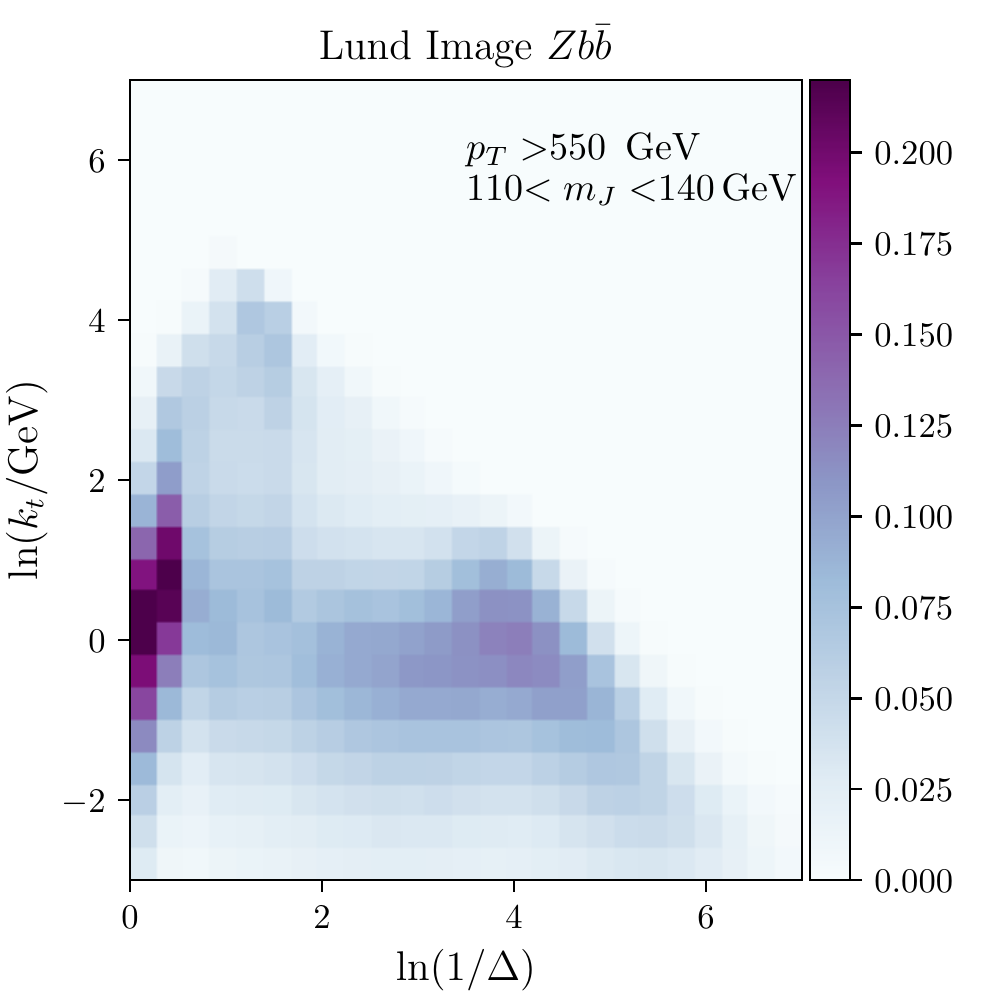}
\includegraphics[scale=0.47]{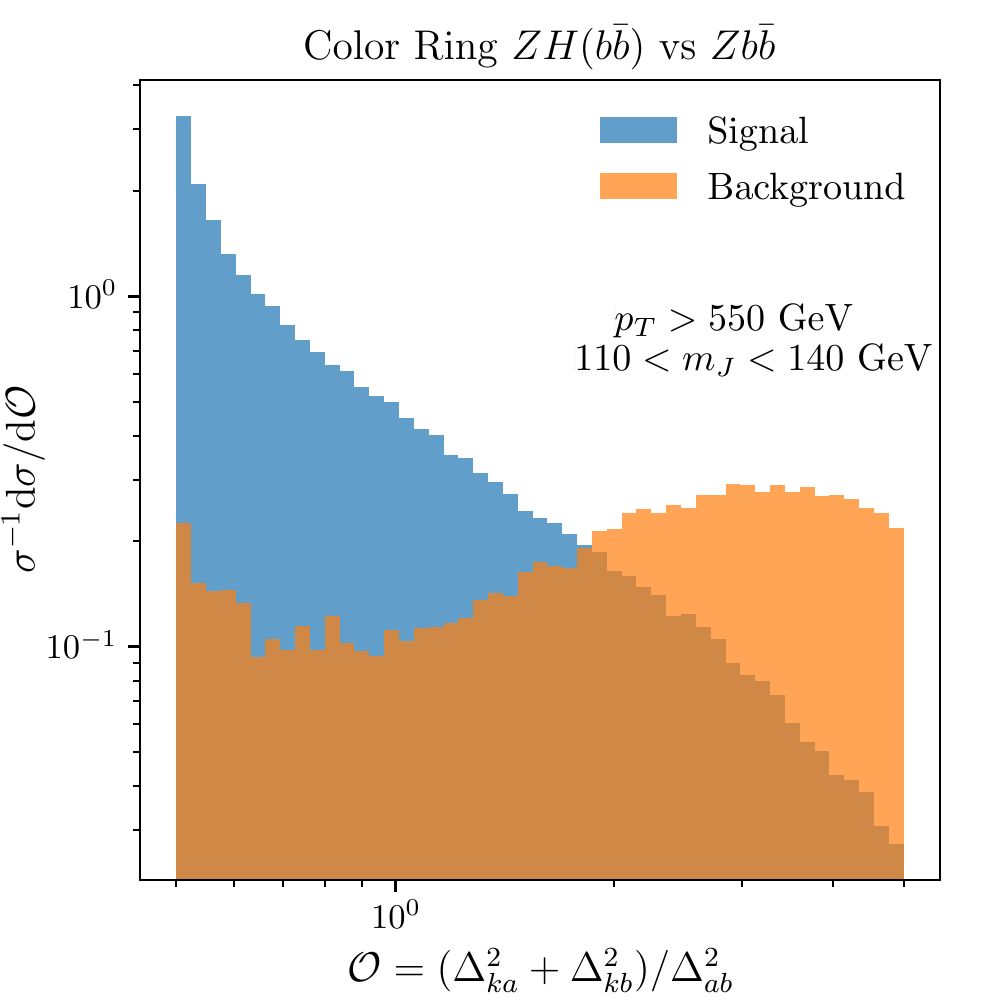}
\caption{The averaged primary Lund jet plane images and the jet color ring distributions for the $p_T>$ 250 GeV and 550 GeV benchmarks of $H \to b \bar b$ analysis. Figures are taken from Ref\cite{Khosa:2021cyk}  under the Creative Commons Attribution 4.0 International license.
}\label{hbbjmc_input}
\end{figure}

\section{H $\to$ $b \bar b$ Analysis}\label{sec:htobb}
\begin{figure*}
\centering
\includegraphics[scale=0.55]{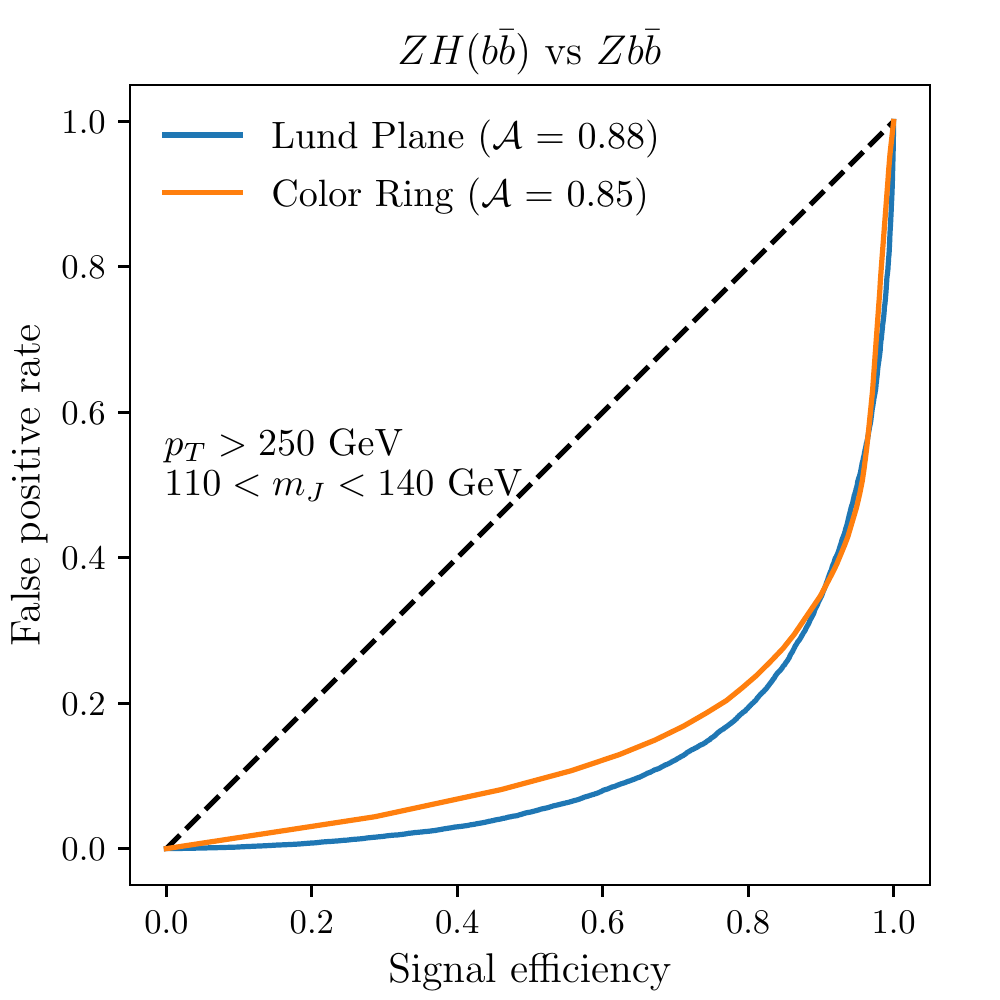}
\includegraphics[scale=0.55]{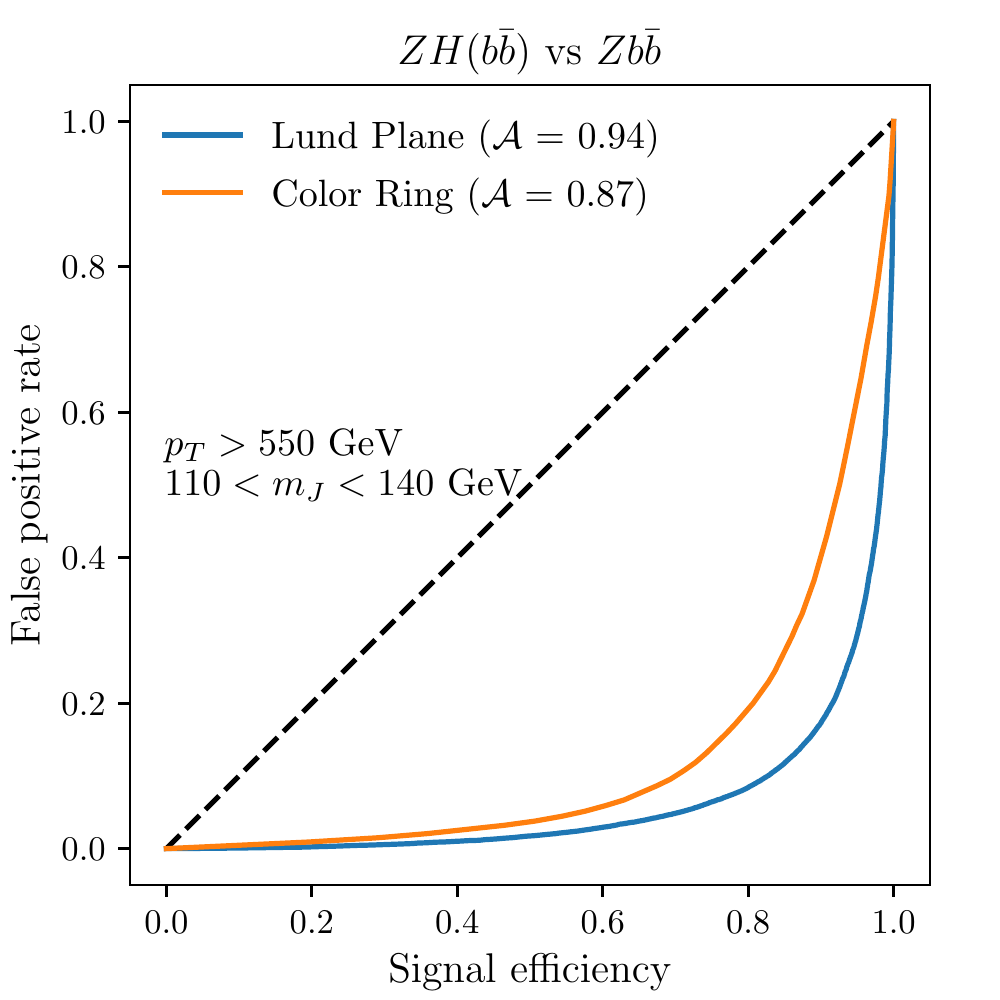}
\caption{CNN (used for the Lund jet plane images) and color ring ROC curves for $H \to b \bar b$ analysis \cite{Khosa:2021cyk}.}\label{hbbjmc_ROCs}
\end{figure*}

Averaged Lund jet plane images and jet color ring distributions of the signal and background are shown in Fig.~\ref{hbbjmc_input} for both the $p_T$ benchmark points (BPs). The main difference in the Lund plane images is in the high $k_T$ and large $\Delta$ region between the two $p_T$ benchmarks. Last column in Fig. \ref{hbbjmc_input} has the jet color ring signal and background distributions.  Signal jet color ring distributions are peaked towards the lower values and background distributions are comparatively flat.  The CNN architecture is presented in Table~\ref{tab:CNNarchi}. These benchmarks are labelled as BP1 ($p_T>$ 250 GeV) and BP2 ($p_T>$ 550 GeV). We compare the Receiver Operating Characteristic (ROC) curves AUC\footnote{AUC is the area under the curve. Note that we use $\mathcal{A}$=1-AUC as a performance measure.} in Fig.~\ref{hbbjmc_ROCs}. We can see CNN classification accuracy for the Lund jet images is slightly better than that of using a cut on the jet color ring (which by construction is an optimal observable to tag a color singlet).

\begin{table}
\centering
\begin{tabular}{lcccccc}
\hline
\hline & BP1 & BP2  & BP3 & BP4 \\
\hline
\hline
$N_1$ Conv2D   & 16    &  16        &  16   &  16      \\
$N_2$ Conv2D   & 16    &  16       &  16   &  16     \\
Dropouts       & 0.25  &  0.05    & 0.20   &  0.20    \\
$N_3$  Conv2D  &  32   &  16        &  16   &  32      \\
$N_4$  Conv2D  & 32    &  16  &  16   &  32      \\
Dropouts       &-      &  0.05   &  0.30  &  0.30      \\
Flat Layer   & 800   & 800       &  800   & 800    \\
Epochs         & 15    & 15      &  20   &  20   \\
Batch Size     & 1000  & 1000     &  800   &  700    \\
\hline
\hline
\end{tabular}
\caption{CNN architectures used for different data sets. Here $N_i, i=$1,..4  represents the number of filters in the $i^{th}$ convolutional layer.}
\label{tab:CNNarchi}
\end{table}

\section{H $\to$ gg Analysis}\label{sec:htogg}
\begin{figure*}
\centering
\includegraphics[scale=0.47]{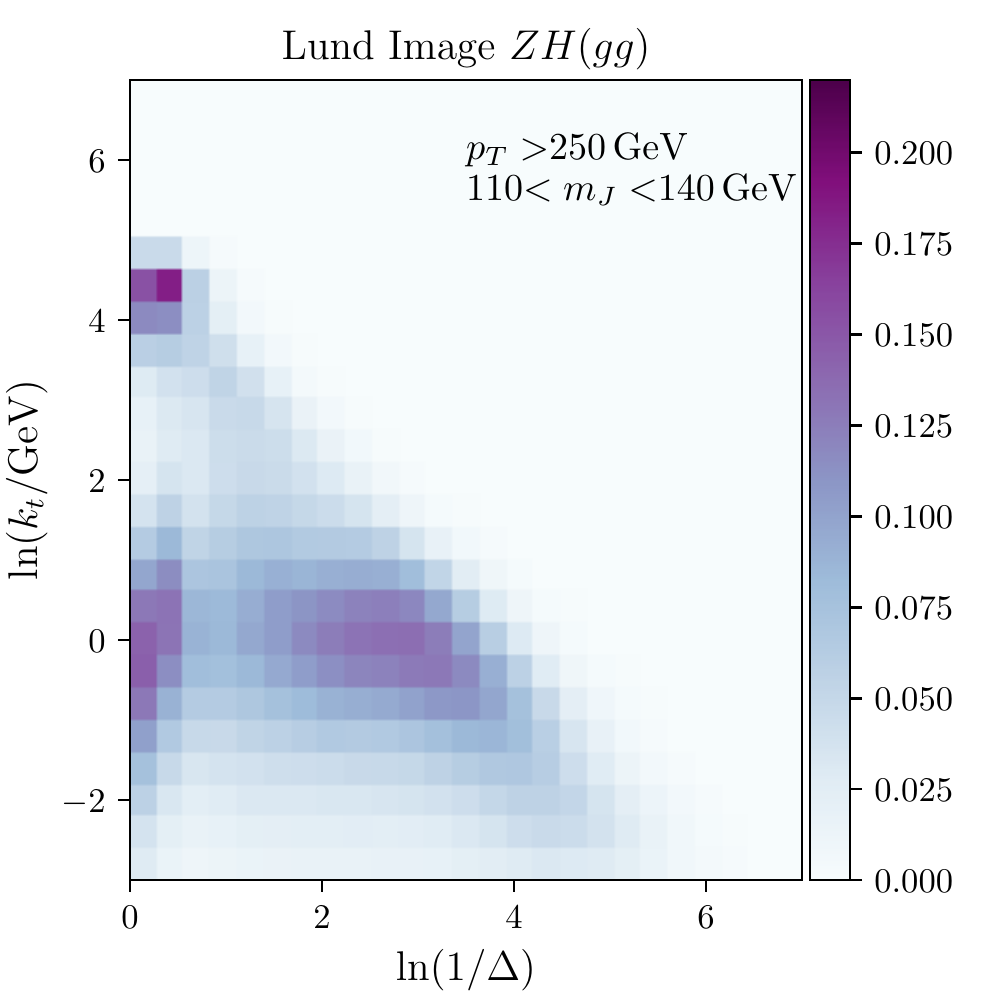}
\includegraphics[scale=0.47]{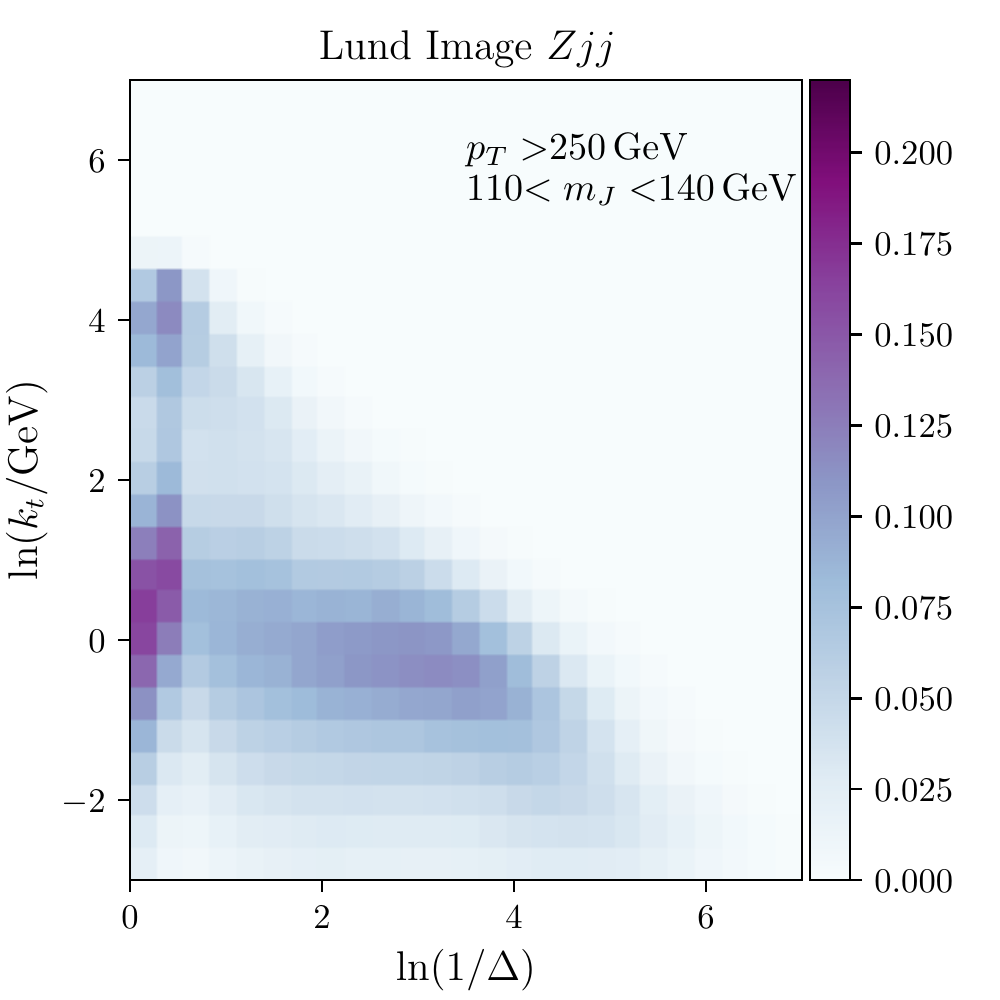}
\includegraphics[scale=0.47]{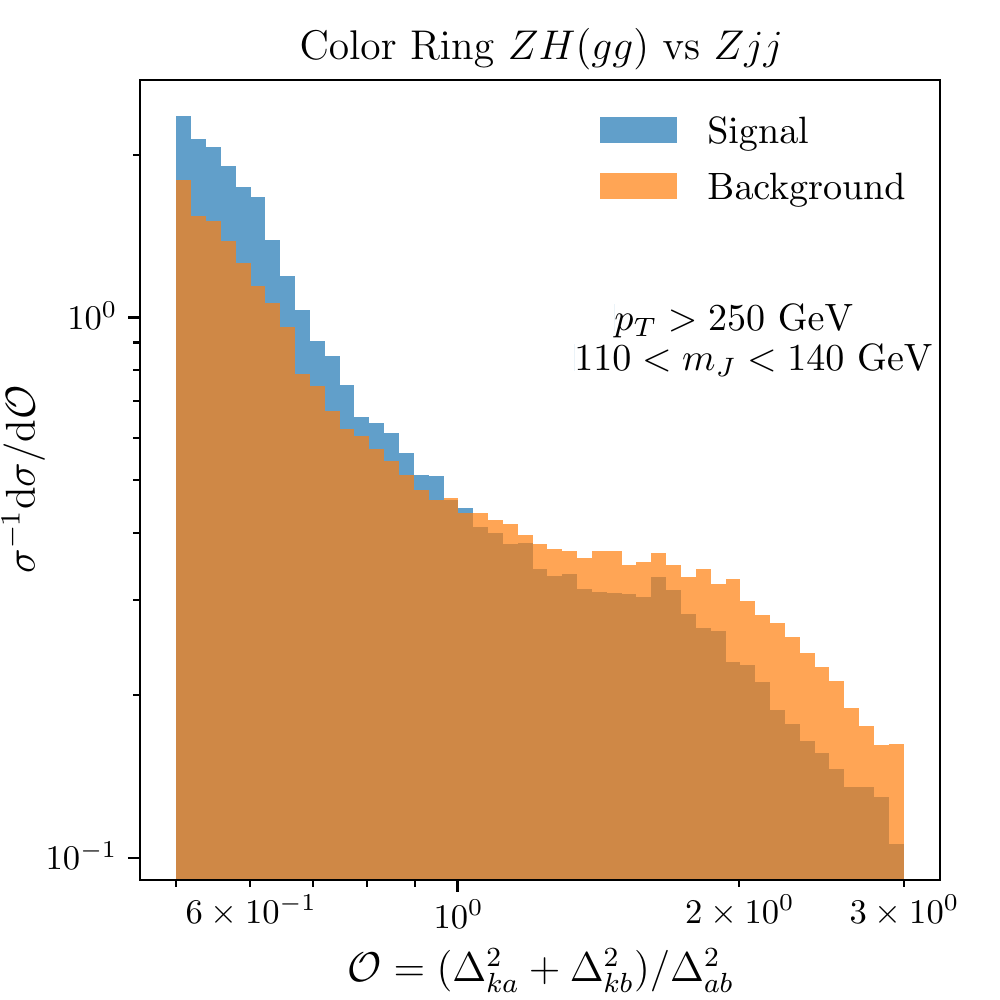}
\includegraphics[scale=0.47]{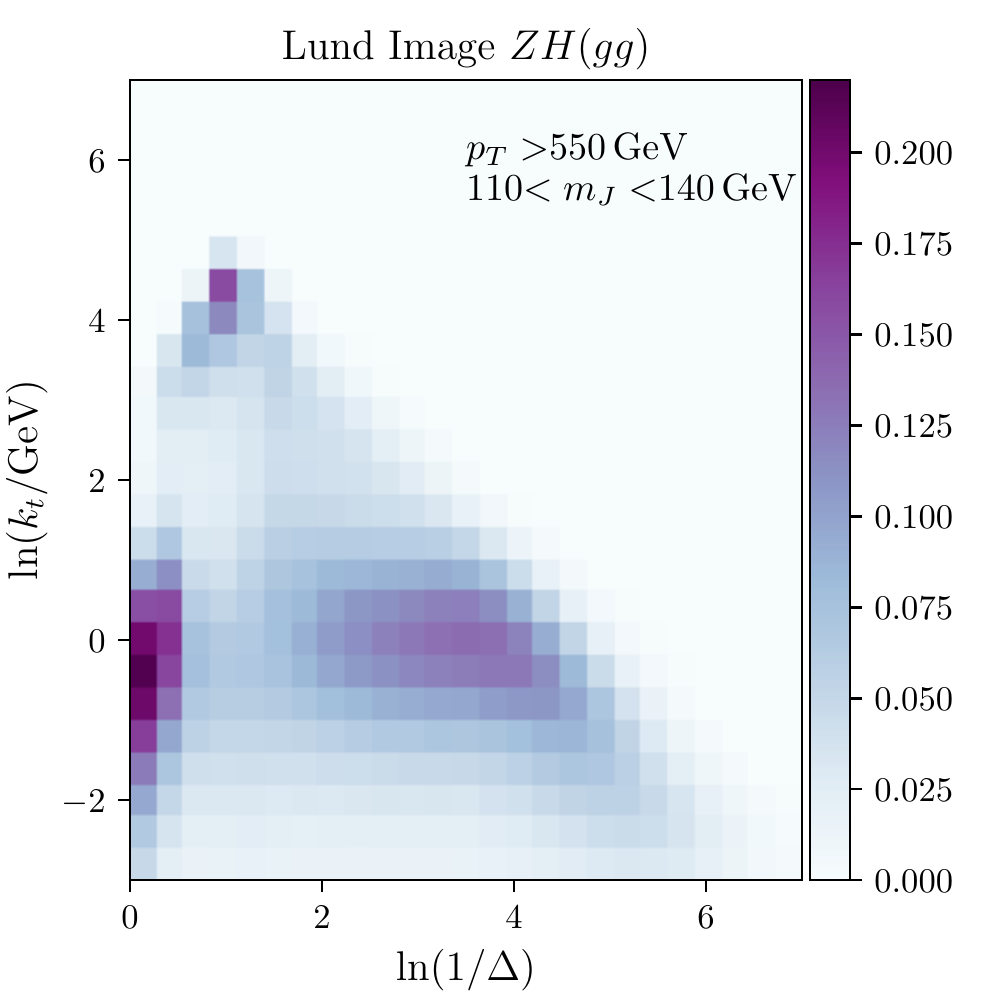}
\includegraphics[scale=0.47]{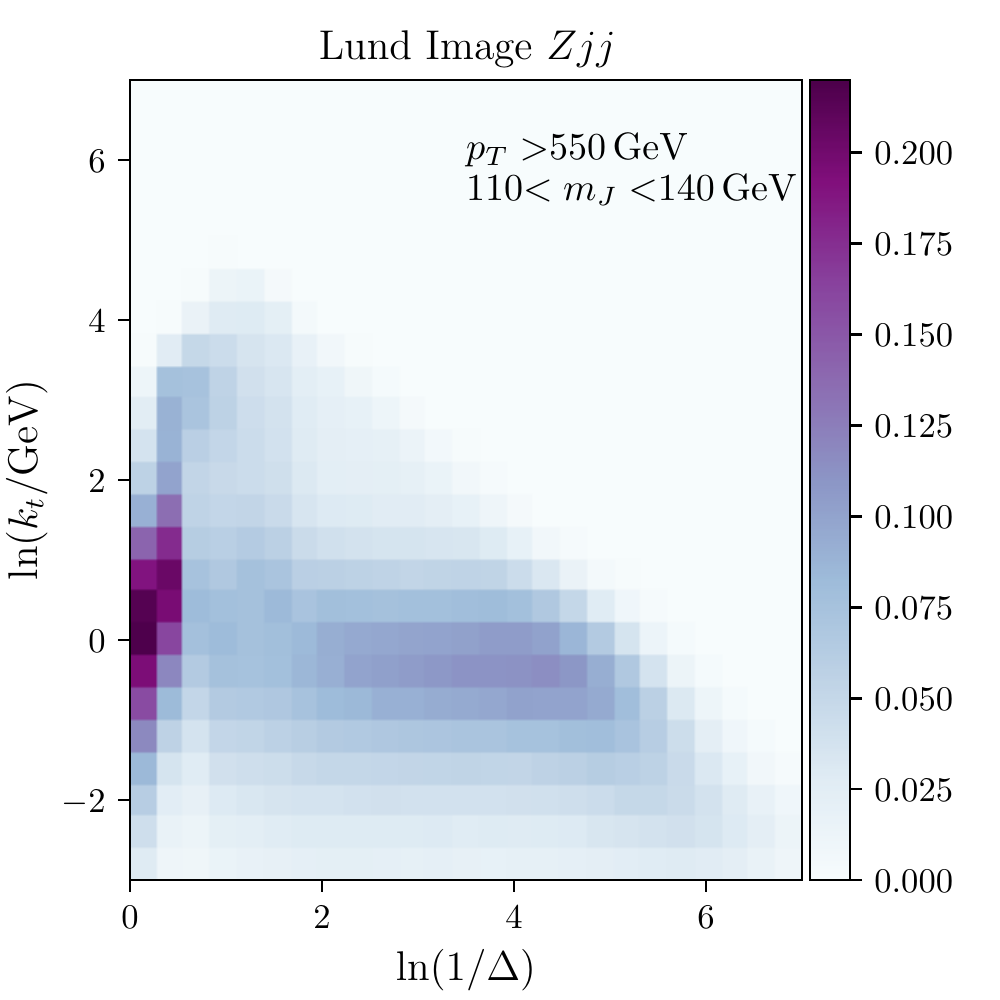}
\includegraphics[scale=0.47]{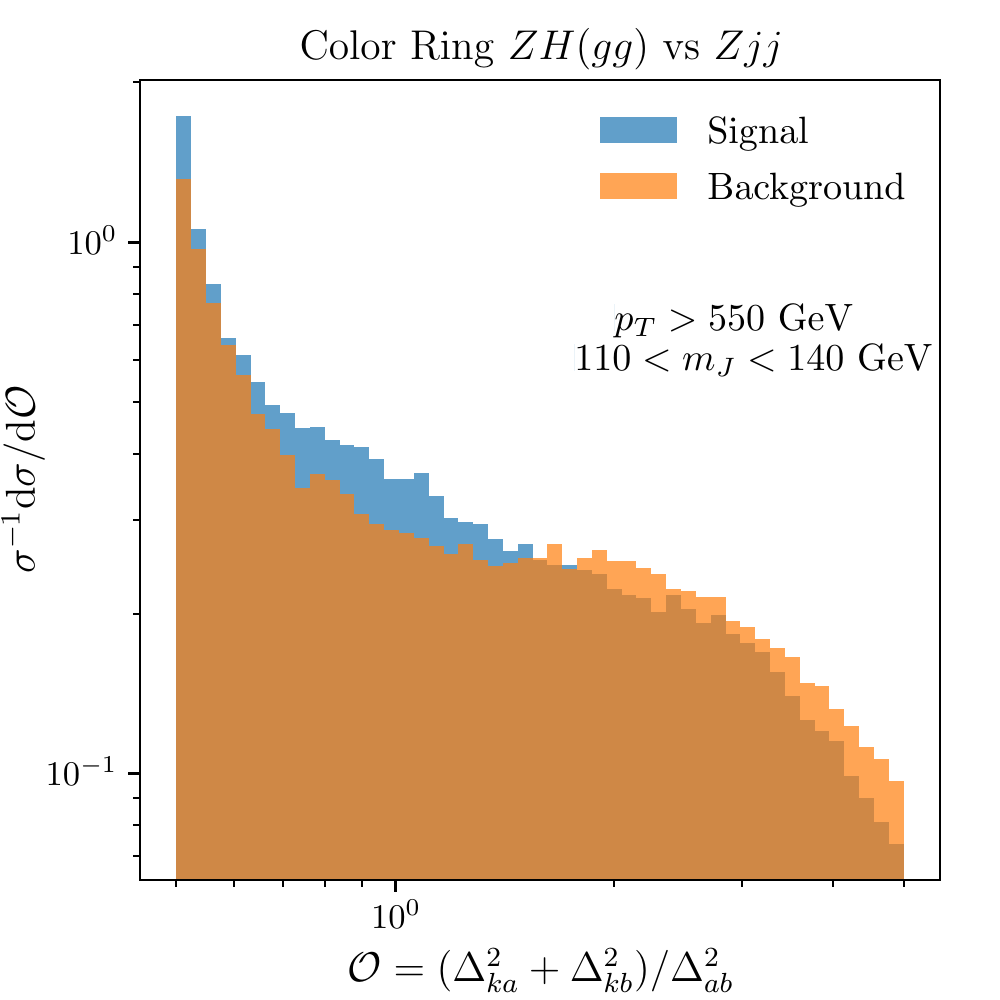}
\caption{The averaged primary Lund jet plane images and the jet color ring distributions for the $p_T>$ 250 GeV and $p_T>$ 550 GeV benchmarks of $H \to gg$ analysis. Figures are taken from Ref\cite{Khosa:2021cyk}.}\label{hggjmc_input}
\end{figure*}

\begin{figure*}
\centering
\includegraphics[scale=0.55]{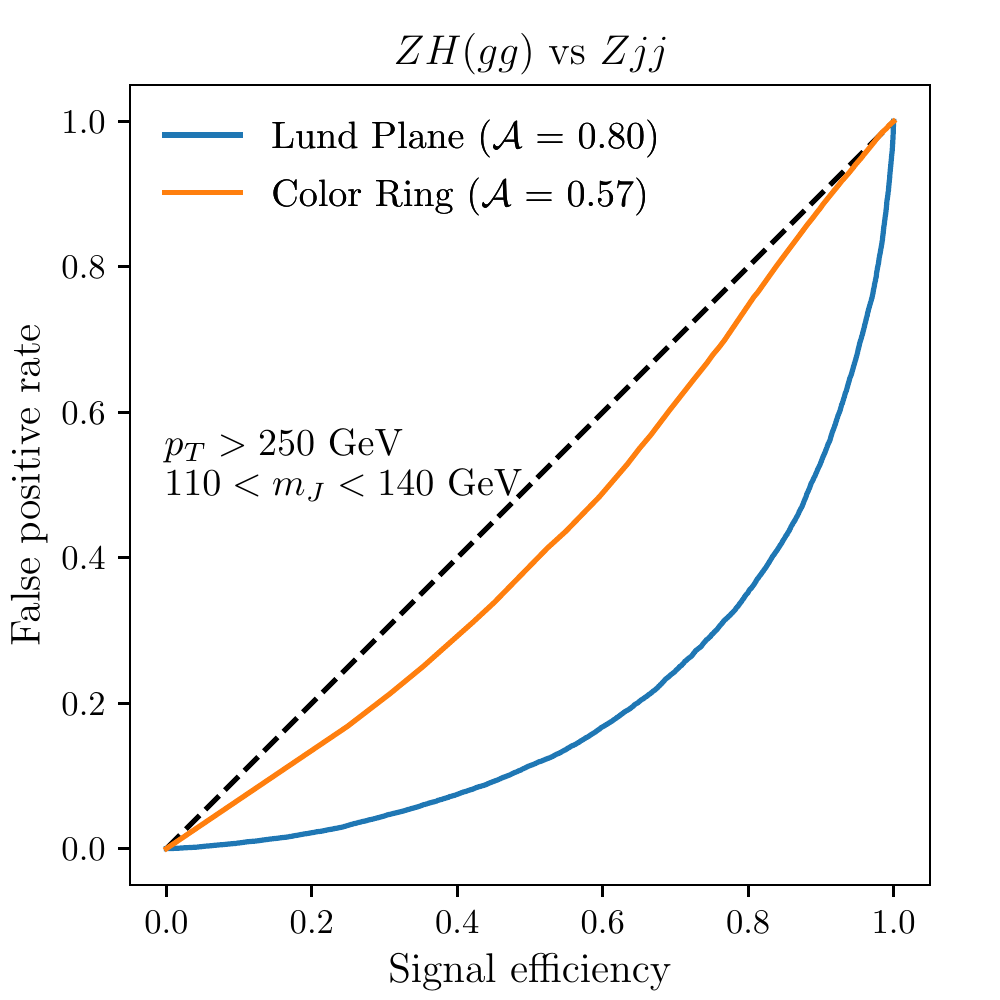}
\includegraphics[scale=0.55]{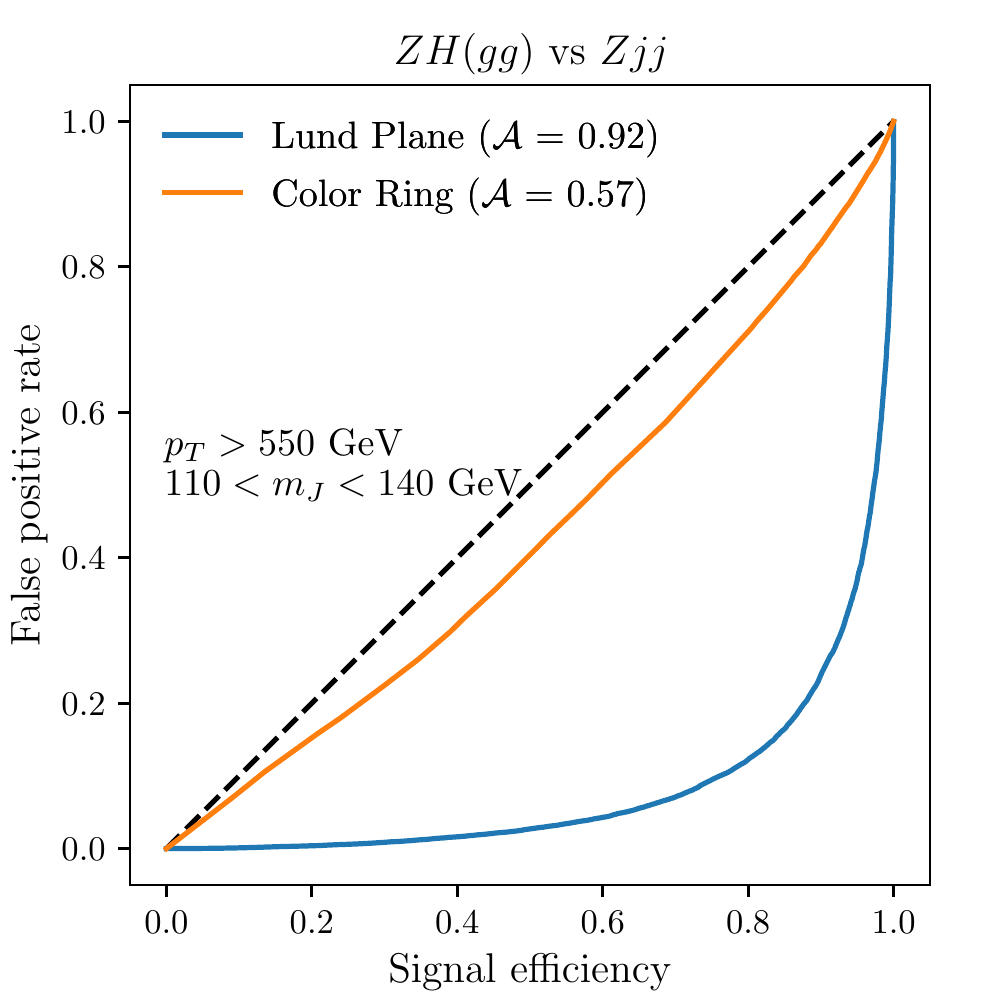}
\caption{CNN (used for the Lund jet plane images) and color ring ROC curves for the $H \to gg$ analysis \cite{Khosa:2021cyk}.}\label{hggjmc_ROCs}
\end{figure*}

 Lund jet plane averaged images and jet color ring distributions are given in Fig.~\ref{hggjmc_input}. We can easily spot the difference between the signal and background Lund jet images however, jet color ring distributions are almost overlapping. Signal distributions are the same as that of $H \to b \bar b$ analysis but the background distributions are very different. In this case several possible color configurations are contributing to the background, contrary to the earlier case where the background was mainly due to $g \to b \bar b$ splitting. As shown in Fig.~\ref{hggjmc_ROCs}, in this case color ring offers no distinction but CNN can  efficiently distinguish the signal and the background. CNN architecture for 250 GeV and 550 GeV benchmark is given by BP3 and BP4 columns in Table~\ref{tab:CNNarchi}.

\section{Conclusions}
We consider the Lund jet plane for Higgs tagging. Using CNN for the signal-background classification, we compare the performance of this set-up with the jet color ring observable. We found the Lund plane and CNN combination provides equally good tagging efficiency for the $H \to b \bar b$ and $H \to gg$ analysis. For the first case, the jet color ring also provides similar tagging performance, but it does not offer any separation for the $H \to gg$ decay. This study further motivates the combined analysis of color sensitive observables and Lund jet plane for the optimal Higgs tagging.

\section*{Acknowledgements}
I thank Simone Marzani for the collaboration on the work presented here. This work is supported by Universit\`a di Genova under the curiosity-driven grant ``Using jets to challenge the Standard Model of particle physics'' and by the Italian Ministry of Research (MUR) under grant PRIN 20172LNEEZ.

\end{document}